# PHOTOCHEMICAL MODEL
# OF IMPACT-PRODUCED LUNAR ATMOSPHERE

A. A. Berezhnoi*, B.A. Klumov**

* Sternberg State Astronomical Institute,
Universitetskij pr., 13,
119899 Moscow, Russia
Tel. 007-095-939-1029/ Fax 007-095-932-8841
email: ber@sai.msu.ru
** Institute of the Geosphere Dynamics
Russian Academy of Sciences
Leninskij prospect, 38/6
117979, Moscow, Russia
Tel. 007-095-939-7989/ Fax: 007-095-137-6511
email: boris@ufn.msk.su

## ABSTRACT

The Lunar Prospector mission detected hydrogen-containing compounds in the lunar polar regions. We have considered the cometary hypothesis of the origin of polar volatiles on the Moon. In our previous paper (Ref. 1) we have found that significant part of the comet material is captured by the lunar gravitation just after the impact. For typical cometary impact the parameters of a such impact-produced atmosphere can be estimated as follows: the concentration of gases near lunar surface is $10^{11}$-$10^{13}$ cm$^{-3}$, the height scale is about 30-100 km. It has been shown that volatile compounds of such artificial atmosphere almost completely delivered into the cold traps if photochemical processes can be discarded (Ref. 1).

In the present paper we use our model (Ref. 1) with photochemical processes to be included to estimate chemical composition of volatiles which can be captured by the cold traps.

## INTRODUCTION

The possible existence of ice in the lunar polar regions was first considered in Ref. 2. The lunar equator is tilted with respect to the ecliptic by approximately 1.5°. For this reason, permanently shadowed areas of the lunar surface exist at the bottom of numerous craters in the polar regions. The surface temperature of these areas is so low that they can be cold traps for water and for a number of other volatile compounds. On the illuminated sections, however, sunlight rapidly dissociates water molecules, and the products of dissociation escape from the Moon's gravity. For this reason, if water does exist on the lunar surface, then it exists only in cold traps and in the form of ice.

Investigation of the radar properties of the lunar surface from the space probe Clementine revealed regions with anomalous radar properties near the lunar south pole (Ref. 3). This apparently attests indirectly to the presence of ice on the Moon. The American spacecraft Lunar Prospector was launched in January 1998. One of its main tasks was to investigate the lunar polar regions by neutron spectroscopy to confirm these results.

The first results of the Lunar Prospector mission were announced in March 1998. Lunar Prospector detected hyrogen-containig compounds in lunar cold traps. If all detected hydrogen is in the form of water ice the volume fraction $\delta_{ice}$ of water ice in a half-meter layer of soil near the surface (deeper layers are inaccessible for measurements by this method) ~ 1.5 %, while the mass of the water ice in the lunar polar regions is estimated to be $3*10^{14}$ g (Ref. 4). What is the origin of hydrogen on the Moon?

The content of volatiles (such as, for example, $H_2O$, $CO_2$, $SO_2$) in a subsurface layer of soil in cold traps is controlled by the following processes: The sources of volatiles are comet and asteroid impacts, micrometeorite bombardment, and solar wind, while losses of these compounds are due to micrometeorite bombardment and the solar wind (Ref. 5).

## COLLISION OF A COMET WITH THE MOON

Let us examine the collision of a comet with the Moon. Typical impact parameters for the Moon are: Comet size $D_{com}$ ~ 2 - 5 km, comet velocity $V_{com}$ ~ 10 - 50 km/s ( most likely velocities $V_{com}$ ~ 20 - 25 km/s), comet density $\rho_{com}$~ 0.3 - 1 g/cm$^3$ , and comet mass $M_{com}$~ $10^{15}$ - $10^{17}$ g. Under such an impact the following occur in succession: essentially complete vaporization of the cometary matter; vaporization, melting, and fragmentation of the regolith by the shock wave from the explosion; ejection of the mixed multiphase cometary matter and lunar soil; and, formation of an impact crater.

The mass $M_{ej}$ of the material ejected on impact can be estimated, using the relation from Ref. 6. In our case ($V_{com}$~ 10 - 20 km/s) one has $M_{ej} / M_{com}$ ~ 10, the mass of the liquid phase ~ 3 $M_i$ , and the mass of the impact vapor ~ $M_{com}$ (Ref. 7).

Some of the matter escapes from the Moon. The escape velocity $V_{esc}$ for the Moon is equal to 2.4 km/s.

Using the velocity distribution of the ejected matter (Ref. 6) it is easily estimated for the impact parameters considered that the mass of the ejected matter with velocities exceeding $V_{esc}$ $M(> V_{esc}) \sim M_{com}$, i.e., mainly hot impact vapor escapes from the Moon's gravity. As the similarity theory shows, the velocity distribution along the radius $r$ in the impact vapor cloud is close to linear: $V(r) \sim V_{max}*r/R_{max}$, where $V_{max}$ is the expansion velocity of the hot cloud and in our case $V_{max} \sim (0.3-0.5)V$. The cloud expands freely: its radius grows with time as $R_{max} \sim V_{max}*t$. The mass of the vapor of cometary origin that remains in the lunar gravitational field, i.e., that has a velocity less than $V_{esc}$ is of the order of $(V_{esc}/V_{max})^3 M_{com} \sim (10^{-3} - 10^{-1})M_{com}$. We note that as the impactor velocity increases, $V_{max}$ ceases to depend on $V_{com}$: $V_{max} \sim (\eta M_{com} V_{com}/M_v)^{0.5}$, where $\eta \sim 0.1$ is the fraction of the impactor kinetic energy that is converted into the impact vapor and $M_v$ is the mass of the impact vapor. For large $V_{com}$ ($V_{com}^2 >> E_v$, where $E_v$ is the heat of vaporization of the soil) $M_v \sim V_{com}^2$ and $V_{max}$ no longer depends on $V_{com}$. For low speed collision the mass of comet origin captured by the lunar gravitation field is in order of 10 % of comet mass.

LUNAR TEMPORARY ATMOSPHERE

In summary, the falling of a comet leads to the formation of an atmosphere on the Moon. We note that condensation processes in the expanding impact vapor can increase the mass of cometary matter on the Moon and, correspondingly, because of the rapid vaporization of the matter by sunlight they can increase the mass of the atmosphere. What is the chemical composition of such an atmosphere?

As the hot cloud expands, quenching occurs: The chemical composition of the impact vapor stops changing, and corresponds to the equilibrium composition at the moment of quenching. The chemical composition of the impact vapor is determined by the composition of the comet and the lunar soil. The fraction of volatiles in the impact vapor, however, decreases as the impactor velocity increases $\sim M_{com}/M_v \sim V_{com}^{-2}$. So impact vapor consists of the cometary matter and the lunar soil, and the ratio of these materials depends on the impact velocity: for example for impacts with velocities $V_{com} \sim 20 - 25$ km/s the ratio is of the order of 1. Thermodynamic calculations show that since the lunar soil does not contain significant quantities of volatiles, the addition of lunar matter to the cometary matter essentially does not change the chemical composition of the volatiles in the impact vapor.

We shall assume that the elemental composition of a comet is identical to that of Halley's comet: H - 48.4, O - 30.4, C -13.7, N - 2.3, Mg - 1.1, Fe - 1.1, S - 1.0, Al - 0.1. (Ref. 8). Then in the wide range of thermodynamic parameters (1000-2000 K, $10^{-2}$ - 10 bar) the main compounds of the impact-produced atmosphere are $H_2$, CO, $H_2O$, $CO_2$, $N_2$, $H_2S$ (see fig. 1) and trace components are $CH_4$, HSOH, SO and $SO_2$. The mass of water in impact vapor is about 5-30 % of comet mass (it depends from quenching temperature and pressure and also from from elemental composition of a comet). We note that sulfur-containing cornpounds will also enter the composition of the impact-produced atmosphere, since at the stage of the explosion of the comet the reaction binding sulfur in the form of iron sulfide is kinetically inhibited.

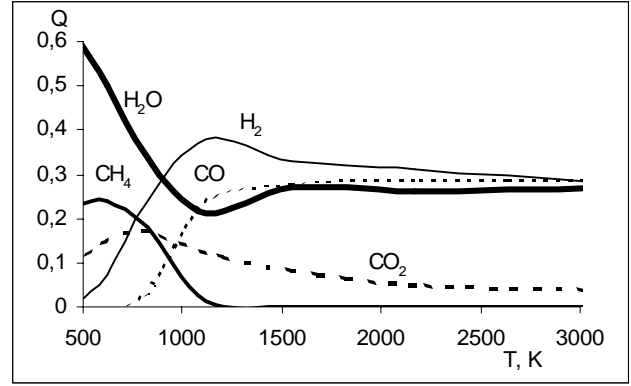

Figure 1. The equilibrium chemical composition of adiabatically cooled fireball of cometary origin. The initial gas temperature is 3000 K, the initial pressure is $10^8$ Pa, $\gamma = 1.2$.

To determine which compounds will condense in cold traps, let us consider the surface temperature $T_c$ of cold traps and the partial pressure of the indicated compounds of the impact atmosphere. The surface temperature at cold traps is less than 100 K (Ref. 9). In the case of the low speed impact of a comet with $D_{com} \sim$ 5 km the mass of the impact-produced atmosphere is $\sim 10^{15}$ g, the pressure at the surface is $\sim 10^{-8}$ bar, the density of neutrals near the surface is $N_a \sim 10^{13}$ cm$^{-3}$. Under these conditions, condensation of $H_2O$, $CO_2$, S, $H_2S$, and $SO_2$ can occur in cold traps.

The lifetime $\tau_a$ of such an atmosphere is determined by the photolysis time $\tau_{ph}$ of the main compounds in the exosphere ($\tau_{ph} \sim 10^6 - 10^7$ s), since the products of photolysis escape the Moon's gravity: $\tau_a \sim \tau_{ph} N_a / N_{ex}$, where $N_{ex}$ is the density of neutrals in the exosphere: $N_{ex} \sim (\sigma H_a)^{-1} \sim 10^8$ cm$^{-3}$, where $\sigma \sim 10^{-15}$ cm$^2$ is the characteristic cross section of elastic collisions for the compounds under study, and $H \sim 50$ km is the scale height of the impact-produced atmosphere. In our case $\tau_a \sim 10^{11}$ s.

The condensable compounds of the atmosphere diffuse into cold traps in a characteristic time $\tau_d \sim (S_m/S_c) H <d_c> N_a \sigma/V$, where $S_m \sim 3*10^{17}$ cm$^2$ and $S_c \sim 3*10^{13}$ cm$^2$ are, respectively, the area of the lunar surface and the area of cold traps (Ref. 4); $<d_c> \sim 0.3$ km/s is the characteristic crater depth; $V \sim 0.3$ km/s is the thermal velocity of gases. So the ratio $\tau_d/\tau_a \sim (S_m/S_c)<d_c>/\tau_{ph}V < 0.1$. Therefore the condensable gases of the impact-produced atmosphere essentially all fall into cold traps.

Because the time of condensation of an atmosphere into lunar cold traps is much greater than the time of photolysis the chemical composition of lunar temporary atmosphere is controlled by photochemical processes. Our kinetic programme consist of 400 gas-phase neutral-neutral reactions with 37 H, O, C, N, S - species: H, $H_2$, C, CH, $CH_2$, $CH_3$, $CH_4$, O, $O_2$, CO, $CO_2$, HCO, $H_2CO$, OH, $H_2O$, $O_2H$, $H_2O_2$, S, $S_2$, CS, SO, $SO_2$, OCS, SH, $H_2S$, N, $N_2$, NH, $NH_2$, $NH_3$, CN, HCN, NO, $NO_2$, HSOH, HSO, $H_2SO$. The rate constants were taken mainly from the UMIST and the CHEMKIN databases. The initial concentrations of indicated species were taken from thermodynamic calculations.

You can see the results of our kinetic accounts on figure 2. It is interesting that quasi-equilibrium chemical composition are reached after $10^7$ s from the moment of creation of such an atmosphere. It easy to see (fig. 3) that the concentration of condensable species ($H_2O$, $CO_2$, $SO_2$) rapidly increases with inscreasing of mass of lunar atmosphere. The quasi-equilibrium chemical composition of temporary lunar atmosphere is sensitive to gas temperature (fig. 4).

So the impact of a small comet with mass less than $10^{14}$ g with low speed (10-15 km/s) or a $10^{16}$ g comet with mean speed doesn't lead to formation of big amounts of ices in lunar cold traps. We note that to explain the amount of ice ($3*10^{14}$ g) observed by Lunar Prospector, a single low speed impact of a comet with $D_{com} \sim 5$ km is sufficient. The Reiner Gamma Formation could be form as a result of a such recent impact (Ref. 10).

## COMPOSITION OF VOLATILES IN LUNAR COLD TRAPS

The chemical composition of ice of cometary origin is determined by the composition of the condensable compounds of the impact-produced atmosphere: The relative concentrations of volatiles in the temporary atmosphere are essentially the same as in the cold traps. We note that the $CO_2$ and $SO_2$ content in the soil is very sensitive to soil temperature, so that the concentrations of these compounds can decrease appreciably over the lifetime of the traps.

The amounts of water observed by Lunar Prospector in cold traps can also be explained by the action of the solar wind and/or micrometeorite bombardment of the lunar surface. However, in these cases the composition of the ice will be different. In meteorite impacts, sulfur and sulfur-containing compounds enter the lunar exosphere. Cold traps capture only sulfur; sulfur-containing compounds are not captured because of the short photolysis time. The solar wind, however, is an inefficient source of sulfur.

Ices of cometary origin can also be distinguished by analyzing the isotopical composition of the ice. An important feature of water of cometary origin is the anomalous high content of deuterium. The ratio D/H in comets is of the order of $3*10^{-4}$ (Ref. 11). Thermodynamic calculations show that as the impact cloud expands, enrichment of the water by deuterium by approximately a factor of 1.5-2 occurs as a result of isotopic exchange. The isotopic composition of hydrogen formed by the interaction of the solar wind with lunar soil is identical to that of the solar wind (D/H< $10^{-4}$), while D/H in water of meteoritic origin corresponds to the isotopic composition of hydrogen in meteorites (D/H~ $(2-5)*10^{-4}$).

## CONCLUSIONS

In summary, the falling of comets onto the Moon leads to formation of ice with definite chemical and isotopic composition in cold traps: The ratio D/H is higher than for other sources of water. Another distinguishing property of "cometary ice" is that it could contain the compounds $SO_2$ and $CO_2$ also.

To explain the amount of ice observed by Lunar Prospector, a single low speed impact of a comet with $D_{com} \sim 5$ km with the Moon is sufficient.

The comet hypothesis of the origin of lunar ice can be checked during the next lunar missions, and if confirmed, then definite progress will have been made in understanding the composition and, accordingly, the nature of comets.

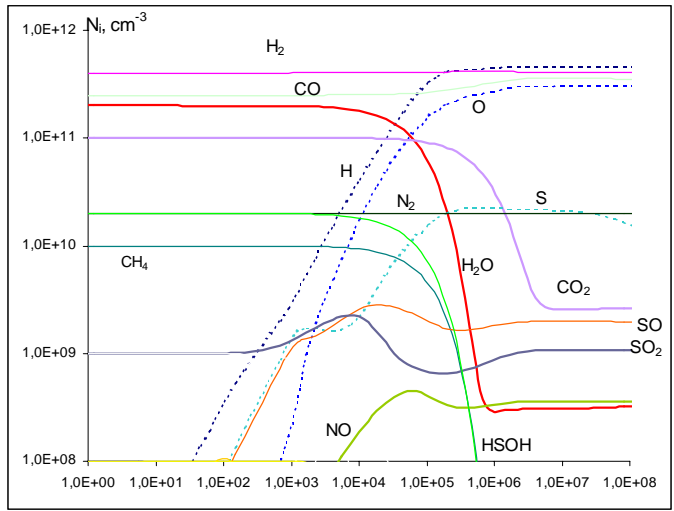

Figure 2. The chemical composition of lunar atmosphere of cometary origin vs time. The gas temperature is 300 K, the number density is $10^{12}$ cm$^{-3}$.

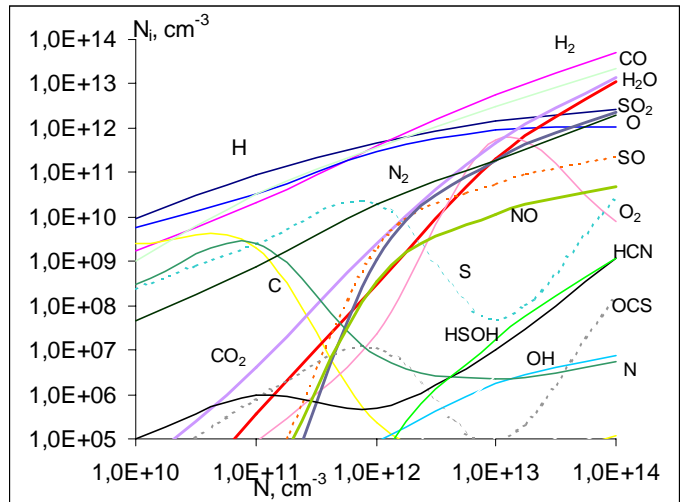

Figure 3. The chemical composition of temporary lunar atmosphere after $10^7$ s from the moment of creation of such an atmosphere vs the total number density. The gas temperature is 300 K.

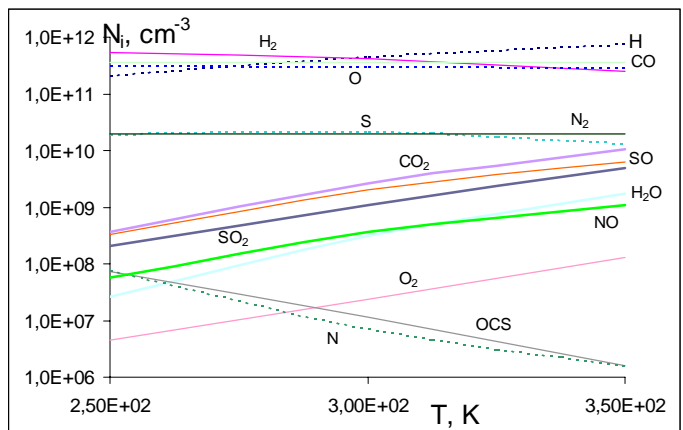

Figure 4. The chemical composition of lunar atmosphere of cometary origin after $10^7$ s from the moment of the creation of such an atmosphere vs the gas temperature. The number density is $10^{12}$ cm$^{-3}$.